This is the accepted manuscript for article “Assessing the Effectiveness of Driver Training Interventions in Improving Safe Engagement with Vehicle Automation Systems”, which is published in Journal of Safety Research, 2025.

# Assessing the Effectiveness of Driver Training Interventions in Improving Safe Engagement with Vehicle Automation Systems

Chengxin Zhang, Ph.D. Candidate
University of Michigan-Dearborn
4901 Evergreen Rd, Dearborn, MI 48128
E-mail: zhangcx@umich.edu

Huizhong Guo, Ph.D.,
University of Michigan Transportation Research Institute
2901 Baxter Rd, Ann Arbor, MI, USA, 48109
E-mail: hzhguo@umich.edu

Zifei Wang, Ph.D. Candidate
University of Michigan-Dearborn
4901 Evergreen Rd, Dearborn, MI 48128
E-mail: zifwang@umich.edu

Fred Feng, Ph.D.,
Industrial and Manufacturing Systems Engineering
University of Michigan-Dearborn
Email: fredfeng@umich.edu

Anuj Pradhan, Ph.D.,
Mechanical and Industrial Engineering
University of Massachusetts
Amherst, MA 01003
Email: anujkpradhan@umass.edu

Shan Bao, Ph.D., Corresponding Author
Industrial and Manufacturing Systems Engineering Department,
University of Michigan-Dearborn, 4901 Evergreen Rd, Dearborn, MI 48128
University of Michigan Transportation Research Institute
2901 Baxter Rd, Ann Arbor, MI, USA, 48109
E-mail: shanbao@umich.edu

## ABSTRACT

**Objective:** This study investigates how targeted training interventions can improve safe driver interaction with vehicle automation (VA) systems, focusing on Adaptive Cruise Control (ACC) and Lane Keeping Assist (LKA), both safety-critical advanced driver assistance systems (ADAS). While VA features can enhance roadway safety, they do not function reliably under all conditions. Drivers' limited awareness of system boundaries can cause overreliance or missed interventions, raising crash risk. Effective training reduces misuse and enhances road safety by promoting correct knowledge and application.

**Method:** A review of multiple automakers' owners' manuals revealed inconsistencies in describing ACC and LKA functions. Three training formats were compared: (1) owners' manual (OM), (2) knowledge-based (KB) with summarized operational guidelines and visual aids, and (3) skill-based hands-on practice in a driving simulator (SIM). Thirty-six participants with no prior VA experience were randomly assigned to one group. Safety-relevant outcomes - system comprehension (quiz scores) and real-world engagement (frequency and duration of activations) - were analyzed using mixed-effects and negative binomial models.

**Results:** KB training produced the greatest improvements in comprehension of system limitations, as well as safer engagement patterns. Compared with OM participants, KB participants achieved significantly higher quiz scores and engaged LKA and ACC more often (1.4 and 1.45 times, respectively); they also demonstrated greater awareness of scenarios requiring manual control, indicating reduced risk of inappropriate reliance. Older drivers exhibited longer activations overall, highlighting age-related differences in reliance and potential safety implications.

**Conclusion:** Short, targeted training can significantly improve safe and effective VA system use, particularly for senior drivers. These results highlight training as a proactive safety intervention to reduce human-automation mismatch and enhance system reliability in real-world driving.

**Practical Applications:** The findings support standardized, accessible, age-tailored training to align driver behavior with VA safety standards, therefore preventing accidents and ensuring safer automation integration for all.

## Introduction

With advances in vehicle sensor technologies and automation algorithms, vehicle automation (VA) systems have become increasingly common in daily driving. These systems aim to reduce driver workload and stress by assisting with critical driving tasks, with the ultimate goal of improving safety. The Society of Automotive Engineers (SAE) defines six levels of driving automation, from Level 0 (no automation) to Level 5 (full automation) (SAE International Standard, 2021). Current deployment is concentrated between Levels 1 and 3, including longitudinal control with Adaptive Cruise Control (ACC), lateral control with Lane Keeping Assist (LKA), and combined systems such as Tesla's Autopilot and General Motors' Super Cruise. Research has demonstrated that Adaptive Cruise Control contributes to road safety by helping drivers maintain appropriate time headways and lowering the risk of unsafe following distances (Piccinini et al., 2014). A recent U.S. government analysis found that vehicles equipped with LKA were 24% less likely to be involved in fatal road-departure crashes compared to similar non-equipped vehicles, highlighting the system's significant contribution to road safety (Scharber, 2024). Evidence also suggests that ADAS has the potential to improve road safety by mitigating common crash types such as intersection, rear-end, and lane-departure accidents, by providing warnings or automating dynamic driving tasks (Masello et al., 2022).

While these systems redefine the driver's role from active operator to supervisory controller (Lee et al., 1994; Lee et al., 2021), the shift introduces new challenges. Lower-level automation has well-documented limitations - for example, lane-keeping systems perform poorly in adverse weather or without road markings - and a mismatch between system capability and user expectation can reduce safety benefits (Parasuraman et al., 1997). Research also shows that when supervising partially automated vehicles, drivers experience vigilance decrements and reduced hazard detection (Greenlee et al., 2019). Recognizing these risks, the National Highway Traffic Safety Administration (NHTSA) has identified driver training as a critical human factors priority (NHTSA, 2013).

Existing studies suggest that training supports calibrated trust in automation, helping drivers anticipate hazards and engage systems safely (Horswill et al., 2013). However, surveys reveal persistent misconceptions about ACC and LKA limitations among vehicle owners (DeGuzman et al., 2021). Overtrust often arises when users are unaware of system boundaries (Dickie et al., 2009), while trust may erode sharply if those limitations are encountered unexpectedly (Beggiato et al., 2013). Although providing accurate information about capabilities and boundaries can improve trust calibration (Beggiato et al., 2015), the effects are often short-lived and do not always translate into safer behavior. Furthermore, trust is dynamic, influenced by drivers' ongoing recognition of system reliability and performance (Rittenberg et al., 2024), yet current training approaches rarely reinforce this evolving relationship.

Taken together, prior research highlights the critical role of training but also reveals gaps: existing efforts are fragmented, inconsistently effective, and often fail to sustain calibrated trust or address behavioral outcomes. These limitations directly motivate the present study, which systematically evaluates and compares different training strategies to identify approaches that better support safe and sustained engagement with vehicle automation.

## Related Work

### Review of Drivers' Training Methods

Research has explored both knowledge-based training, such as reading owners' manuals or viewing videos/animations, and skill-based training, including driving simulators with intelligent tutoring systems (ITS) that provide real-time feedback (Xiao et al., 2010). While such approaches are foundational, most early ACC studies focused primarily on system descriptions and broad developmental perspectives rather than evaluating drivers' learning processes or limitations in real-world use.

Evidence also indicates that training can improve safety-relevant skills, but limitations remain. Horswill et al. (2013) demonstrated that even highly experienced drivers exhibited suboptimal hazard perception and required targeted training to improve performance, suggesting that experience alone does not ensure adequate risk awareness. Similarly, Beggiato et al. (2015) showed that drivers developed trust and acceptance of ACC with repeated exposure, but also found that unexperienced system limitations tended to disappear from users' mental models—highlighting the risk of overreliance without reinforcement of boundaries. Studies comparing training formats, such as manuals versus multimedia, revealed limited improvements in knowledge and virtually no change in driver behavior (Nobel et al., 2019), pointing to the difficulty of translating instructional content into safe practice. Furthermore, Singer et al. (2024) demonstrated that training emphasizing system capabilities rather than limitations fostered overconfidence, even when accurate information was provided.

Overall, prior work underscores the importance of training but reveals persistent gaps: many approaches lack long-term effectiveness, fail to reinforce awareness of system boundaries, and risk unintentionally fostering overconfidence. These limitations justify the need for research into alternative training strategies that combine efficient delivery with a clear, safety-focused emphasis on system limitations. This study directly addresses this gap by systematically comparing different training formats and evaluating both knowledge outcomes and behavioral measures of safe automation use.

### Review of Owners' Manual

A review of owners' manuals (OMs) from 12 automakers was conducted to explore how ACC and LKA are described, with a focus on system capabilities and safety considerations. Manuals were collected from online sources (e.g., https://mycardoeswhat.org) and examined for terminology, function descriptions, and safety guidance.

The analysis showed opportunities to enhance clarity and consistency across manuals, which could further support safe driver interaction with automation. Key observations include:

- Terminology Differences: ACC is sometimes referred to as Dynamic Radar Cruise Control, Distronic Plus, or Smart Cruise Control, while LKA is described as LaneSense or DISTRONIC PLUS with Steering Assist. Establishing more uniform terminology could help drivers more easily recognize and apply their knowledge across different vehicle brands.

- Function Descriptions: Activation thresholds vary, with ACC reported as operating between 25–45 mph and LKA between 32–90 mph. Providing harmonized operational ranges may give drivers clearer expectations and support safer use.
- Safety Guidance: Manuals often highlight conditions such as radar interference, reflections, or adverse weather (e.g., heavy fog) where systems may not perform optimally. Offering more detailed situational guidance would further empower drivers to make informed decisions and remain prepared to intervene when needed.

Overall, these findings suggest that standardized and accessible communication strategies could strengthen drivers' understanding of automation features, encouraging both confident use and safe readiness to take control when necessary.

To understand current training practices for new vehicle buyers, the research team visited local dealerships and reviewed online resources. Findings indicate that training on automated features is not yet consistently integrated into the delivery process. A J.D. Power report notes that most dealership time is spent on vehicle selection and negotiation (about 60 minutes each), followed by 30 minutes for paperwork and 30 minutes for delivery. Within this limited timeframe, incorporating comprehensive training on vehicle automation can be challenging (J.D. Power, 2015).

Recognizing this gap highlights opportunities to design training that is both efficient and tailored to drivers' diverse preferences and driving styles. Identifying strategies that positively influence driver engagement - particularly for middle-aged and older drivers - will be essential for supporting safe and confident use of automation. Future work can also expand training approaches to account for individual differences such as age, experience, and gender, ensuring that automated features are introduced in ways that maximize safety, trust, and usability.

The primary objective of this study is to evaluate the effectiveness of three training methods in enhancing middle-aged and senior drivers' safe operation and safety-oriented trust calibration when using Level 2 and Level 3 automated in-vehicle systems. The study focuses on two safety-relevant functions - ACC and LKA - and examines both knowledge-based and skill-based training strategies. By supporting accurate mental model development and long-term trust calibration, the study aims to promote safe engagement with automation and ensure drivers remain prepared to intervene when necessary. This work addresses the following safety-focused research questions:

1) What training strategies for VA systems are currently available, and how effectively do they promote safe driver understanding, based on literature, owners' manuals, and dealership practices?
2) What types of training content and delivery formats most effectively communicate the safe operation and limitations of VA functions?
3) Among owners' manuals, knowledge-based training, and skill-based training, which method best improves drivers' safety-related understanding and on-road performance?
4) How do different training methods, together with individual user differences, influence safety outcomes such as proper system use, driver readiness to intervene, and balanced trust in automation?

## Method

The primary objective of this study is to examine the effects of different training methods on drivers' trust calibration and their safe interaction with vehicle automation technologies. To support this objective, a comprehensive review of owners' manuals from multiple automakers was conducted to identify consistencies and inconsistencies in the descriptions of ACC and LKA functionalities and limitations. Drawing from this review, eight representatives 'corner cases' were developed to illustrate critical scenarios associated with these systems (see **Table 1** and **Table 2**). Training materials were then created using consistent content centered on the operational features and limitations of ACC and LKA but delivered in varied formats according to the assigned training condition. Examples of these materials are provided in Figures 1 through 3, with additional details outlined in the Training Material Preparation section.

Based on the information collected from literature review, knowledge-based (tablet-based text/graph) and skill-based (driving simulator-based hands-on) training materials were created to address the safe operation and limitations of the Level 2 VA features, LKA and ACC. Next, scientific experiments were carried out to compare the effectiveness of three training methods: owners' manual only, knowledge-based training, and skill-based training. A mixed-subject experiment with lab and on-road sessions was designed and conducted with 36 participants. Finally, the relationship between training methods and outcomes (system understanding and engagement) was examined using mixed regression models and negative binomial models. This structured approach allowed the research to systematically evaluate training methods while incorporating a blend of theoretical and practical assessments, ultimately enhancing our understanding of how to better prepare drivers for the effective use of VA systems. This study received approval from the Institutional Review Board at the University of Michigan.

## Training Material Preparation

### *Materials for Owners' Manual (OM) Training*

Owners' manuals (OMs) were collected from various online sources, including official manufacturer websites and educational platforms (e.g., https://mycardoeswhat.org). Manuals from different automakers were provided to participants assigned to the OM training group. These participants reviewed information directly related to selected VA systems - namely LKA and ACC - including system symbols, descriptions, and operational instructions. **Figure 1** presents sample training materials extracted from the OMs for both ACC and LKA systems.

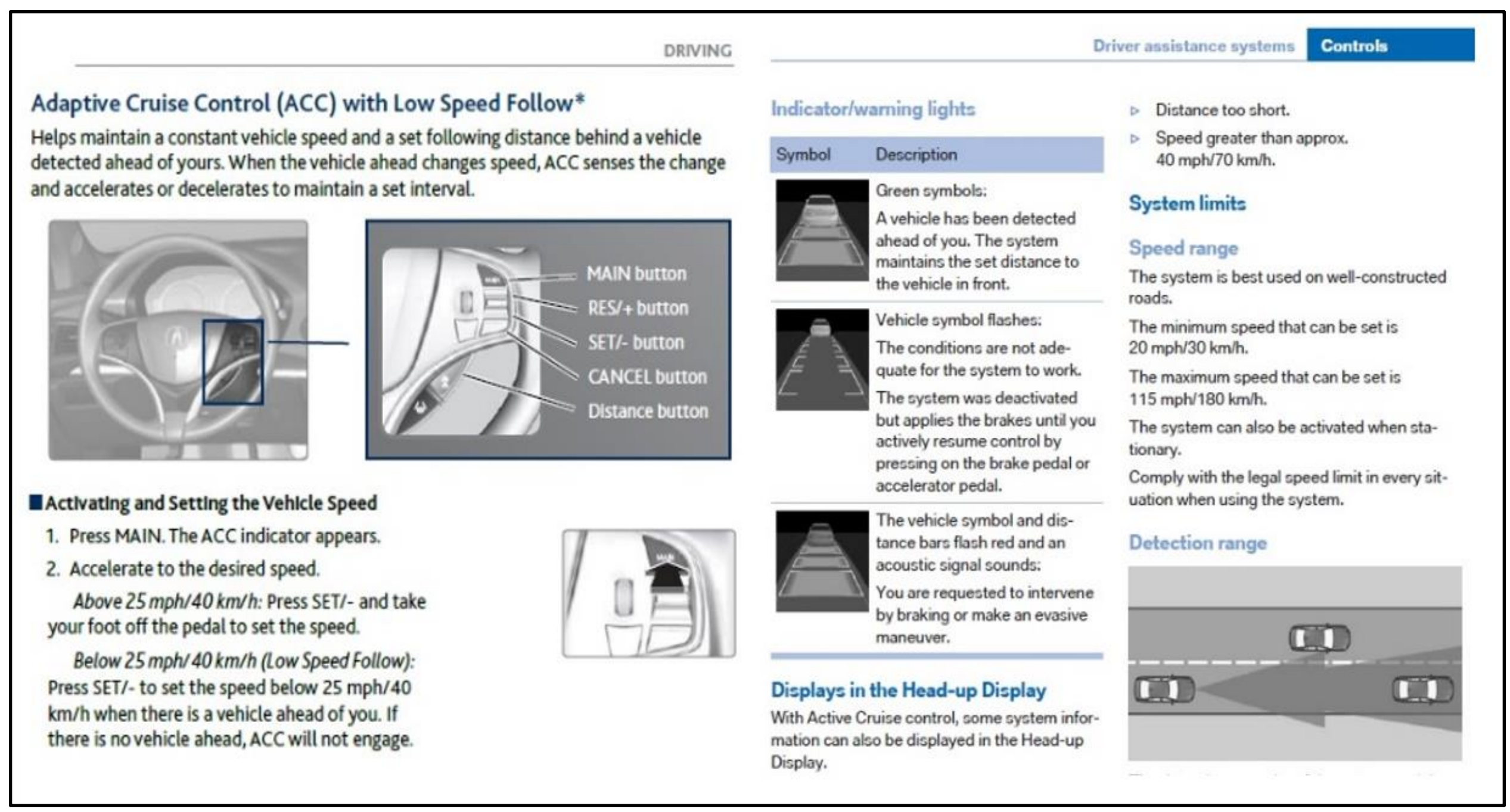

**Figure 1. Sample of Owners' manuals (OM) training material**

### *Materials for Knowledge-based Training*

Knowledge-based (KB) text training materials on the functionality and limitations of ACC and LKA were developed based on a comprehensive review of various owners' manuals. These materials were designed to provide clear and concise information to enhance middle-aged and senior drivers' understanding and proper usage of these automated systems. To address the potential lack of emphasis on system limitations, the training content was designed to include summarized descriptions, operational guidelines, and visual aids that clearly conveyed both the capabilities and boundaries of the VA systems. Supplementary text and graphical materials were incorporated to illustrate scenarios in which system failures or limitations may occur, and to support drivers' understanding of the system's operational design domain, including instructions on when manual takeover is required to ensure safer and more effective human-vehicle interaction. Figure 2 illustrates a sample of the KB tablet-based training materials, which incorporate text and graphical representations to facilitate learning and retention.

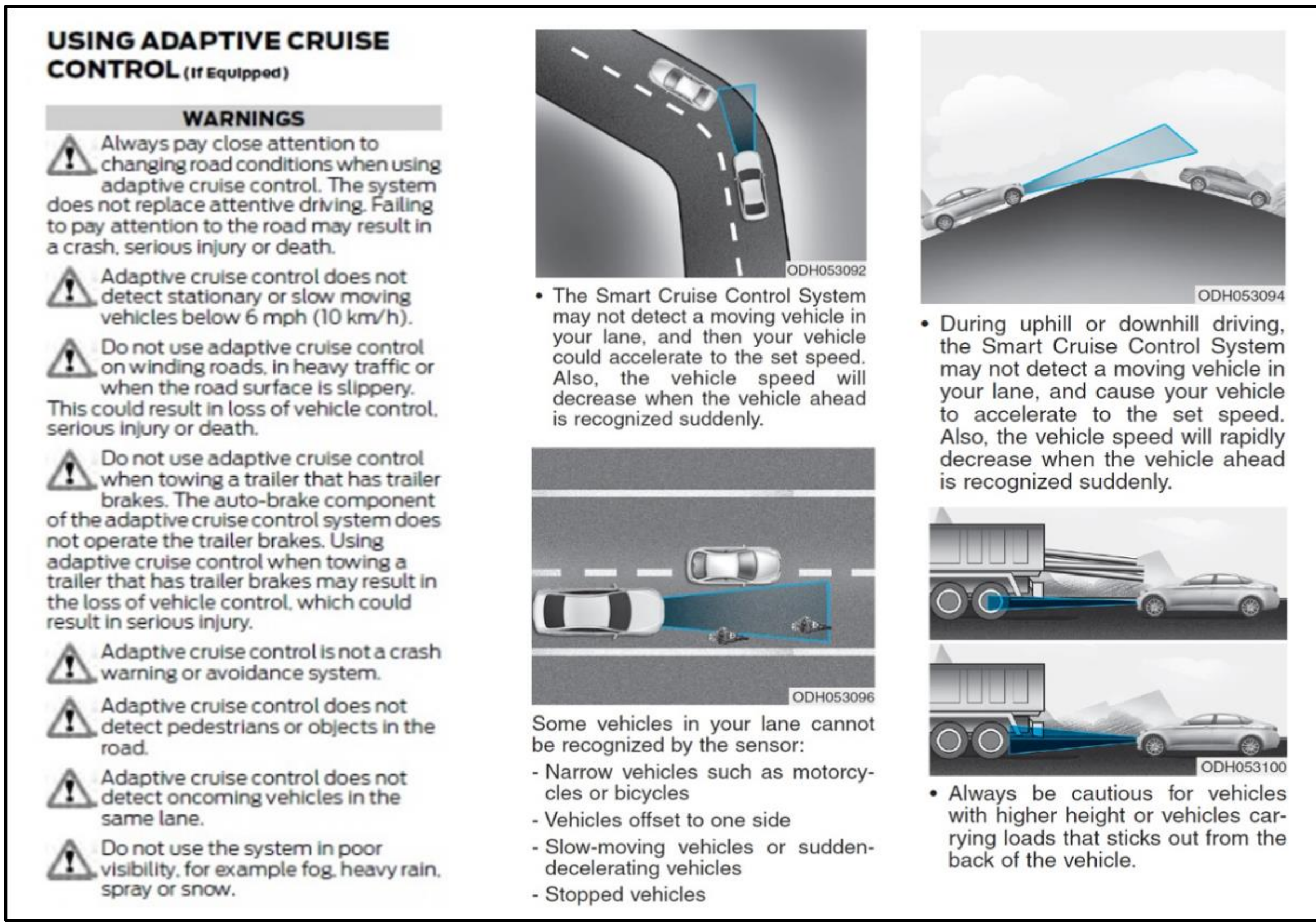

**USING ADAPTIVE CRUISE CONTROL** (If Equipped)

**WARNINGS**

Always pay close attention to changing road conditions when using adaptive cruise control. The system does not replace attentive driving. Failing to pay attention to the road may result in a crash, serious injury or death.

Adaptive cruise control does not detect stationary or slow moving vehicles below 6 mph (10 km/h).

Do not use adaptive cruise control on winding roads, in heavy traffic or when the road surface is slippery. This could result in loss of vehicle control, serious injury or death.

Do not use adaptive cruise control when towing a trailer that has trailer brakes. The auto-brake component of the adaptive cruise control system does not operate the trailer brakes. Using adaptive cruise control when towing a trailer that has trailer brakes may result in the loss of vehicle control, which could result in serious injury.

Adaptive cruise control is not a crash warning or avoidance system.

Adaptive cruise control does not detect pedestrians or objects in the road.

Adaptive cruise control does not detect oncoming vehicles in the same lane.

Do not use the system in poor visibility, for example fog, heavy rain, spray or snow.

ODH053092

- The Smart Cruise Control System may not detect a moving vehicle in your lane, and then your vehicle could accelerate to the set speed. Also, the vehicle speed will decrease when the vehicle ahead is recognized suddenly.

ODH053096

Some vehicles in your lane cannot be recognized by the sensor:

- Narrow vehicles such as motorcycles or bicycles
- Vehicles offset to one side
- Slow-moving vehicles or sudden-decelerating vehicles
- Stopped vehicles

ODH053094

- During uphill or downhill driving, the Smart Cruise Control System may not detect a moving vehicle in your lane, and cause your vehicle to accelerate to the set speed. Also, the vehicle speed will rapidly decrease when the vehicle ahead is recognized suddenly.

ODH053100

- Always be cautious for vehicles with higher height or vehicles carrying loads that sticks out from the back of the vehicle.

**Figure 2. Sample of Knowledge-based (KB) training material**

***Materials for Skill-based Simulator Training***

Through the review of the literature and owners' manuals, a list of corner cases describing the limitations of ACC and LKA systems under various conditions was identified. In this context, corner cases refer to situations where ADAS sensors and systems reach their performance limits and have a high potential to fail. Such conditions may be caused by poor weather, faded or missing lane markings, dynamic interactions with other road users, or complex traffic environments. Drivers often lack awareness of these edge-case system limitations, particularly those that are rarely encountered in daily driving. These situations highlight the importance of training and interface design that explicitly reinforce awareness of system boundaries, ensuring that drivers remain prepared to respond effectively when automation cannot perform as expected. Because some of these conditions are hazardous to test in the real world, simulation provides a valuable opportunity to replicate risky scenarios and examine driver responses safely.

Based on this analysis, eight corner-case training scenarios were developed to address these specific limitations (see **Table 1** and **Table 2**). These scenarios are designed to provide practical, real-world examples to help drivers better understand situations where ACC and LKA systems may not perform as expected. **Figure 3** illustrates one of the skill-based training (SIM) scenarios, along with detailed descriptions of its functionalities and limitations. This scenario, like the others, aims to equip drivers with critical knowledge to ensure they can safely navigate situations where reliance on vehicle automation alone might be insufficient.

**Table 1. Training Scenario List for ACC**

| No. | Name | High-level learning objective | Script |
|---|---|---|---|
| 1 | Motorcycle | ACC may not detect small objects. | The ACC has limited detection zone and capability. It may not detect small-size vehicles immediately in front of you, such as motorcycles, scooters, or bicycles, significantly when offset from the centerline of the lane. If this occurs, you may need to manually control a proper distance from the vehicle. |
| 2 | Cutting-in | ACC has a limited detection zone. | The ACC has a limited detection zone. It may not detect a vehicle entering the lane ahead from an adjacent lane until it has completely moved into your lane. Always stay alert to the traffic and intervene if needed |
| 3 | Hard- braking | ACC has limited braking capacity. | The ACC has limited braking capacity. If the car in front of you brakes hard or quickly stops, the system may be unable to stop your vehicle in time to avoid a crash. Always stay alert to the traffic and be prepared to apply the brake pedal if needed. |
| 4 | Stationary vehicle | ACC may not detect all vehicles. | The ACC may not detect stationary or slow-moving vehicles. If you are approaching a stopped or slow-moving vehicle, the system may be unable to stop your vehicle to avoid a crash. Always stay alert to the traffic and be prepared to apply the brake pedal if needed. |

**Table 2. Training Scenario List for LKA**

| No. | Name | High-level learning objective | Script |
|---|---|---|---|
| 5 | Fog | LKA may not work if the camera cannot see the lane. | The LKA relies on cameras to see lane markings. If the visibility is poor, such as during fog, heavy rain, snow, or other bad weather, the camera may not be able to see the lane markings. Therefore, the system may not work correctly and may be automatically deactivated. Always pay attention to the road condition and be prepared to steer the vehicle if needed. |
| 6 | Lane-closure | LKA may need help understanding the meanings of certain road elements (e.g., traffic cones). | The LKA may not be able to detect lane closure due to road repair indicated by traffic cones or other objects. If the lane is closed ahead, the system may continue to follow the lane markings of the closed segment. Always pay attention to the road condition and be prepared to intervene. |
| 7 | Sharp-curve | LKA has limited capacity in steering the vehicle. | The LKA is not intended for use on sharp curves. The system may not work correctly, and the vehicle may go off-road on sharp curves. Always pay attention to the road condition and be prepared to steer the vehicle if needed. |

| 8 | Faded-lane | LKA may not work if the lanes are not present or well-marked. | The LKA relies on cameras to see lane markings. If the lane markings fade or disappear on the road ahead, the system may not work correctly and be automatically deactivated. Always pay attention to the road condition and be prepared to steer the vehicle if needed. |
|---|---|---|---|

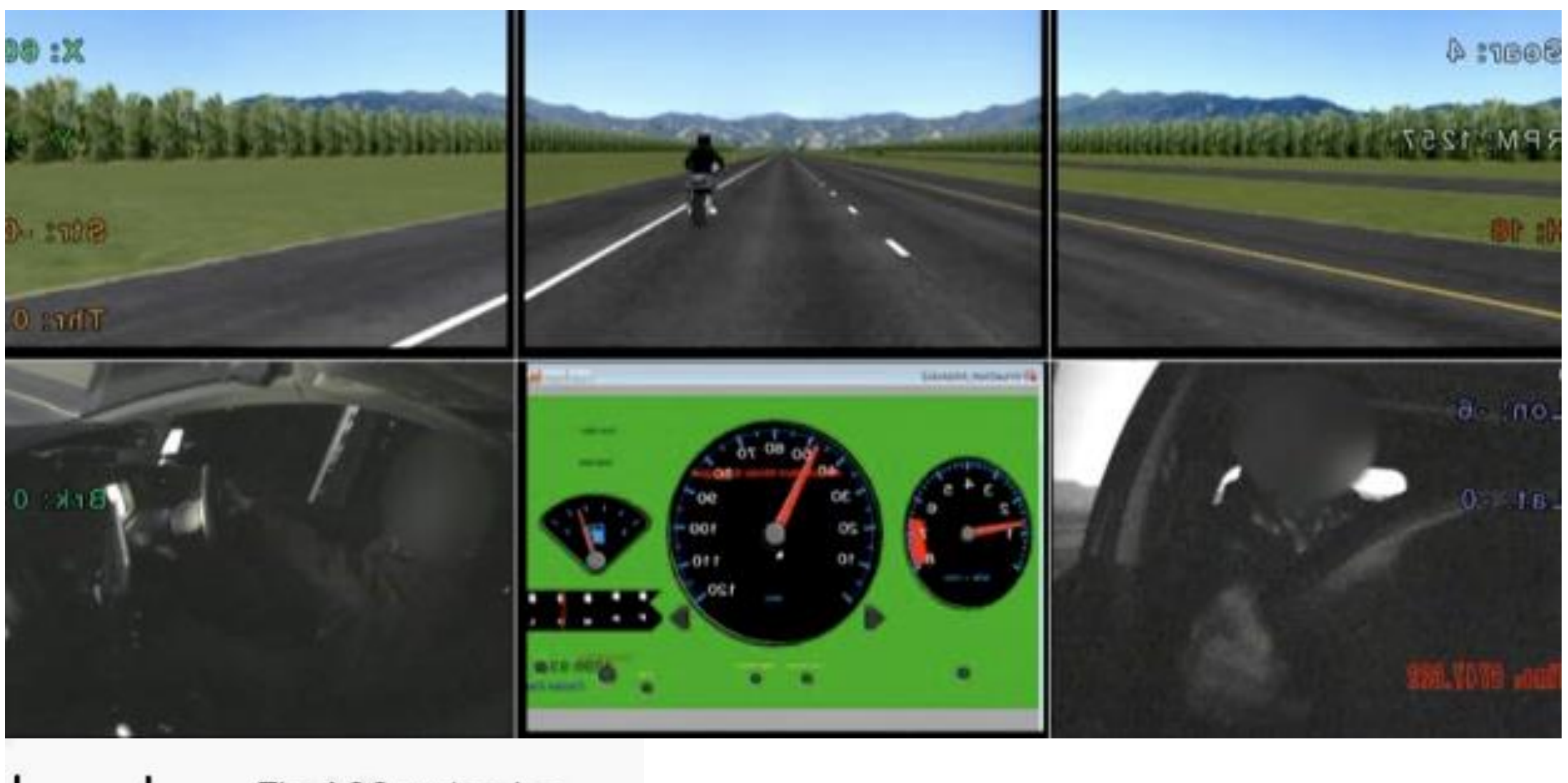

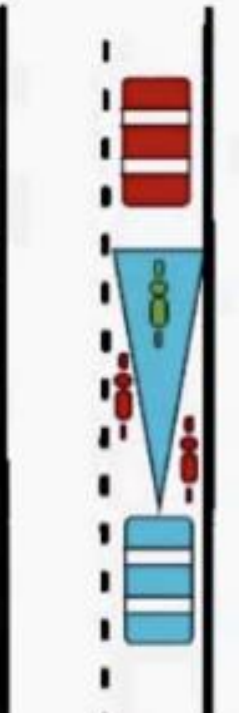

- The ACC system has limited detection zone and capability;
- It may not detect small size ahead, such as motorcycles, scooters, or bicycles;
- Need manually take over control.

**Figure 3. Sample of simulator training (SIM) corner-case scenario and descriptions**

## Design of Experiment

A mixed-subject experiment, comprising both lab and on-road sessions, was designed and conducted with 36 participants. These participants were divided into two age groups (middle-aged and older adults) and balanced by gender to ensure a diverse and representative sample. All participants had no prior experience with the ACC and LKA systems, and each signed a consent form before data collection commenced. Participants were recruited and then randomly assigned in equal numbers to one of three training groups: OM, KB, and SIM. Although the study included three groups of 12 participants, which may present limitations in statistical power, this design was chosen due to practical constraints related to recruitment capacity and study timeline. To mitigate potential underpowering, repeated measures and within-subject comparisons were incorporated to enhance the robustness and internal validity of the findings.

Figure 4 outlines the procedures for the experiment and details the sequential steps taken during both the lab and on-road sessions. The aim was to assess the efficacy of each training method in enhancing drivers' understanding and engagement with the VA systems, both immediately after training and over an extended period.

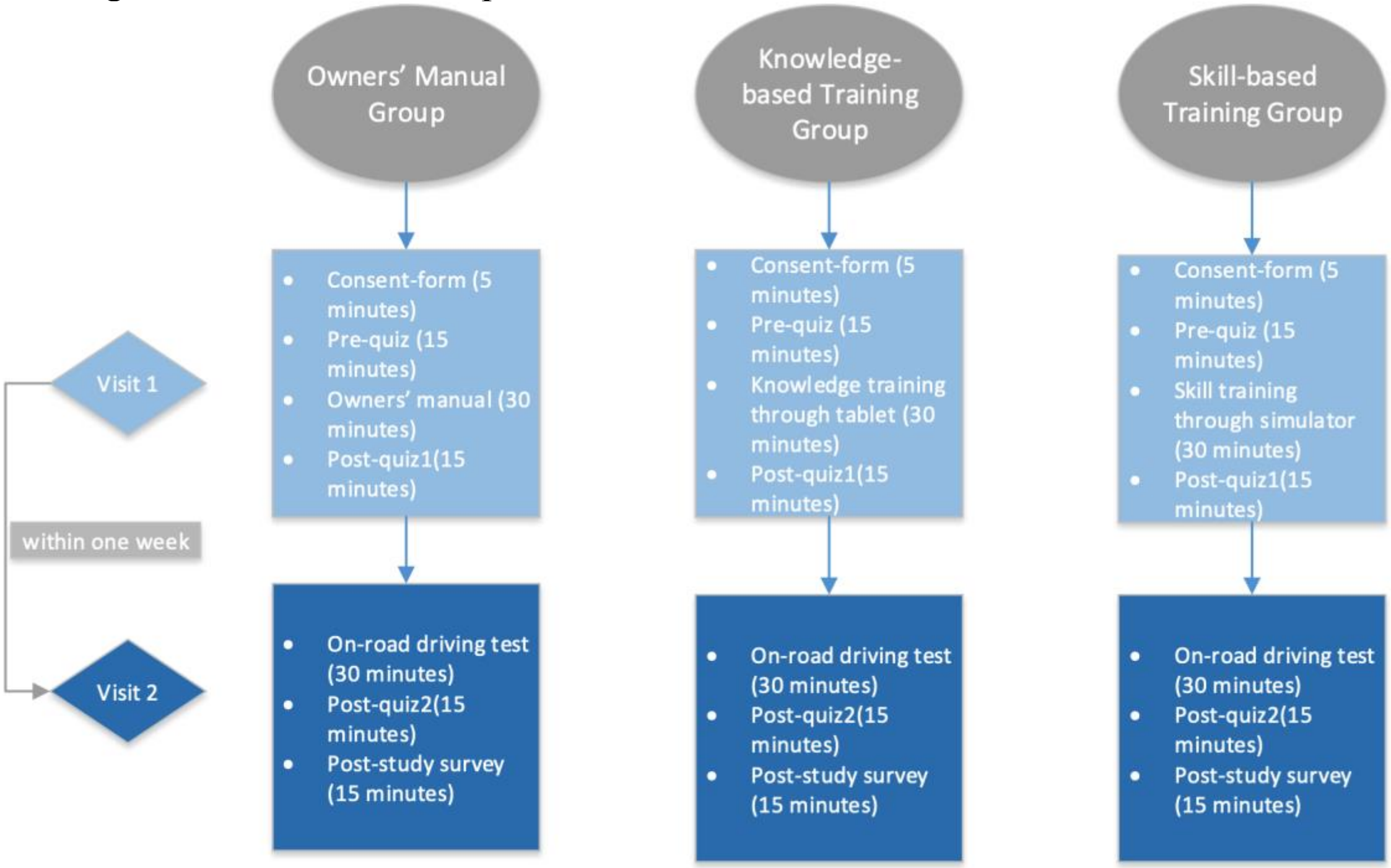

**Figure 4. Flowchart on experimental design**

Each participant visited the University of Michigan Transportation Research Institute (UMTRI) twice, as illustrated in Figure 4. During the first visit, they completed a consent form and received a lab-based training intervention. The OM group was provided with an original owners' manual. The KB group reviewed the designed content using a tablet. The SIM group engaged with eight simulated training scenarios on a driving simulator at UMTRI.

All participants were required to return to UMTRI within one week for an on-road test and a follow-up knowledge check. A questionnaire consisting of both single- and multiple-choice items assessing participants' understanding of ACC and LKA functionalities and limitations was developed and administered to each participant at three separate time points: *before the training, immediately after the training,* and *during a delayed post-training session* (immediately after the on-road drive). The number of correct answers was scored for each participant to quantify the training effect.

For the on-road test, a vehicle equipped with actual driving assistance features - ACC for speed control and LKA for steering support—was rented (Figure 5). During the test, the vehicle was instrumented with two GoPro cameras to record both the dashboard and the driver's actions (Figure 5). The on-road test lasted between 30 to 40 minutes, with a mean duration of 36.7 minutes, and primarily involved highway driving. An experimenter accompanied each driver and informed them that they could be considered an online help source for the VA systems. This was necessary to ensure that participants had access to immediate assistance in case of confusion or technical issues, thereby maintaining safety and ensuring the continuity of the experiment. Due to video quality issues, the video from one subject was excluded, leaving data from 35 subjects for the final analysis.

After the test, video coding was performed using Datavyu (v1.5.3) software. LKA and ACC system engagement frequency were counted. LKA and ACC active duration (continuous) were calculated. Engagement frequency was calculated as the number of times the subjects pressed the button to activate the function (ACC or LKA). Engagement duration was measured as the total time the function (ACC or LKA) remained active, starting from when the subjects pressed the button to activate the function until the function became inactive (function status under standby mode was not included in the engagement duration). For ACC, the function was considered active (engaged) when the speed limiter and distance/time interval symbol appeared on the dashboard. For LKA, the function was active (engaged) when the steering wheel symbol on the dashboard turned green.

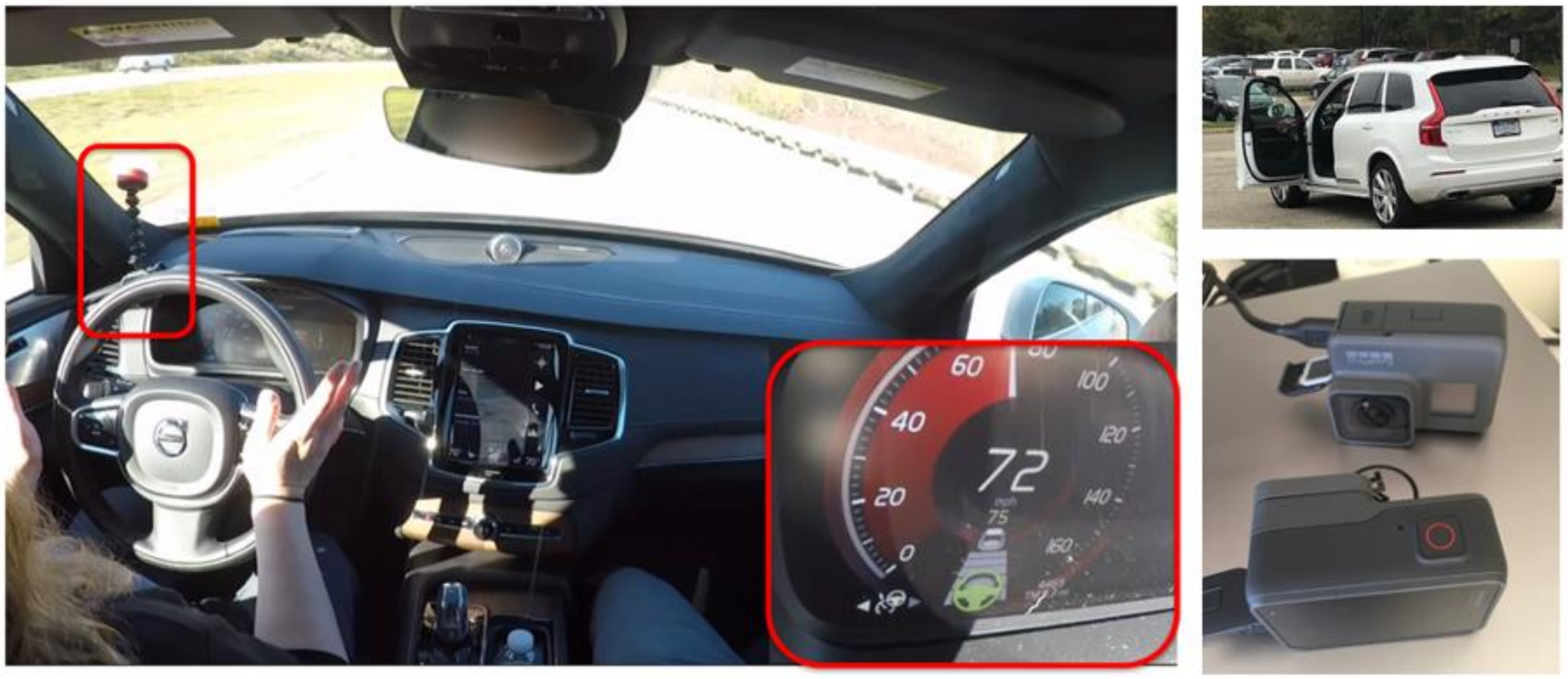

**Figure 5. On-road driving test example**

**Data Analysis - Statistical Modeling**

Two types of data -continuous and count data- were collected from the experiment and used in the analysis: (1) knowledge check scores (continuous), including true/false and multiple choices results, and (2) on-road video-coded data, including the engagement frequency (count) and duration (continuous).

Questionnaire scores were measured three times for each subject: before the training (pre), immediately after the training (after), and immediately after the on-road test (post). Similarly, each participant engaged the ACC or LKA multiple times during the on-road test, resulting in multiple records of engagement duration per participant. To accommodate the repeated measures for each participant, mixed-effects models were employed, as shown in Equation 1 (implemented using Python for data science):

$$y = X\beta + Zu + \varepsilon , \qquad (1)$$

where X is a vector of fixed-effects variables; Z is the random effects, in our context, on the subject; and å is the error term. The mixed-effects model accounts for subject-specific intercepts and/or slopes, allowing for a more accurate representation of repeated measures data by capturing individual differences while simultaneously estimating the effects of the training methods (Fitzmaurice et al., 2008).

Negative binomial regression was used to model the engagement frequency. This model is an extension of Poisson regression, but it relaxes the assumption that the variance is equal to the mean of the outcome. A negative binomial model can be used to model the frequency of event occurrences, making it suitable for frequency analysis (Abdel-Aty et al., 2000).
Based on the data distributions and attributes, mixed regression models were created for score data and engagement duration analysis, while negative binomial models were employed for function engagement frequency data. In total, six models were built to address six distinct outcome variables, as follows:

1) Model 1 – A mixed regression model for true/false scores
2) Model 2 – A mixed regression model for multiple choice scores
3) Model 3 – A negative binomial model for LKA engagement frequency
4) Model 4 – A mixed regression model for LKA engagement duration
5) Model 5 – A negative binomial model for ACC engagement frequency
6) Model 6 – A mixed regression model for ACC engagement duration

True/false and multiple-choice scores were measured as accuracy percentages, ranging from 0% to 100%. The independent variables are listed in Table 3. It should be noted that the variable "test time," which indicates whether the questionnaire was taken before (pre), immediately after the training (after), or after the on-road test (post), is applicable only to the score models (Models 1 and 2). Similarly, "LKA engaged start time," which measures the time into a drive when an LKA engagement was initiated, is relevant only to Model 4. All modeling analyses were conducted using RStudio (Version 2022.12.0+353). Variables with a p-value less than 0.05 were considered significant predictors. Additionally, effect sizes were calculated following Cohen's (1988) recommendations. Cohen's d measures the standardized differences between the means of different groups. The effect sizes were classified as small ($d \geq .2$), medium ($d \geq .5$), and large ($d \geq .8$) (Luger et al., 2023).

**Table 3. Variables for Questionnaire Score Models**

| Variable | Type | Recorded level |
|---|---|---|
| Training Group | Categorical | OM * (owners' manual)<br>KB (knowledge-based)<br>SIM (skill-based simulator) |
| Test time | Categorical | Pre * (score before training)<br>After (score right after training)<br>Post (score after one week) |

| Gender | Categorical | Male<br>Female * |
|---|---|---|
| Age Group | Categorical | Middle *<br>Older |
| LKA engaged start time | Continuous | A continuous value |

(**Note**: base levels are noted by “*”; Pre: score before training; After: score right after training; Post: score after one week; OM: Owners’ manual training, KB: Knowledge-based training, SIM: Skill-based simulator training.)

## Results

### Score Analysis – Mixed-effects Model (Model 1 & Model 2)

As shown in Figure 6, both the true/false and multiple-choice scores increased notably after the training intervention, especially for the true/false questions. However, the scores across the three training groups were similar, with overlapping error bars.

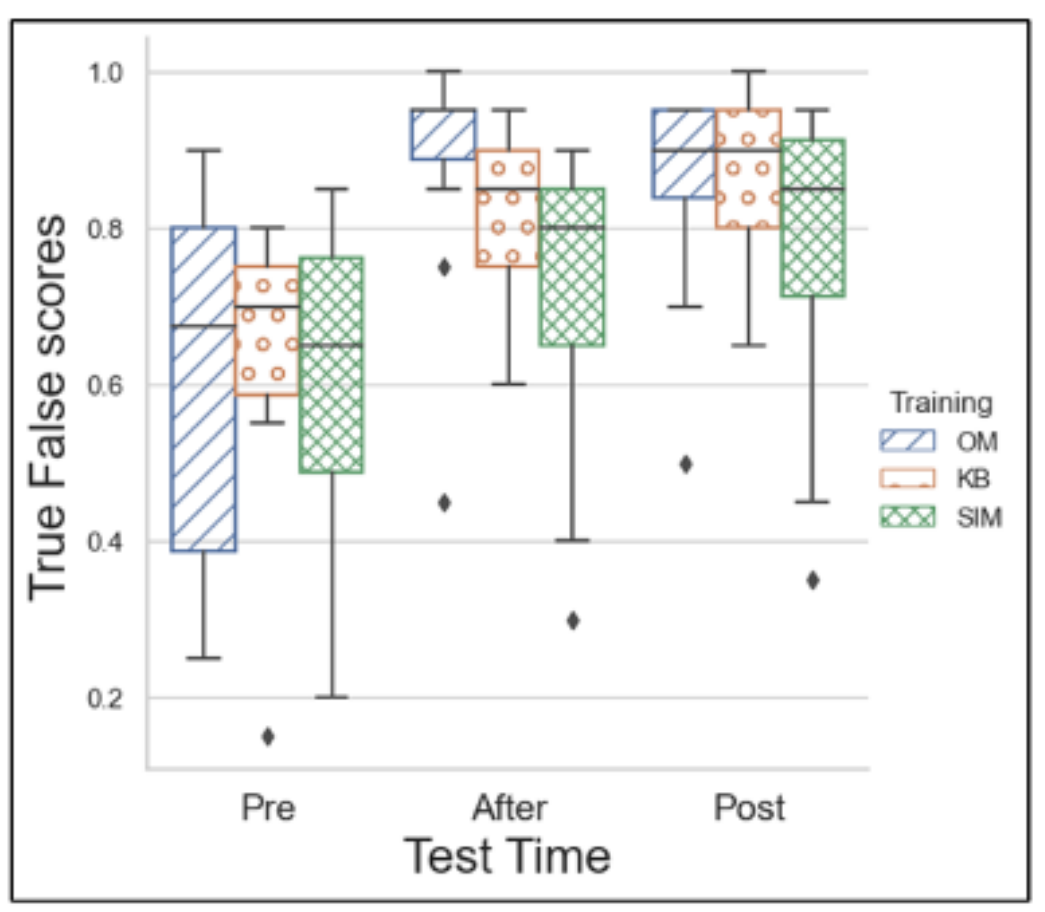

**Figure 6a: True False Score by Test Time**

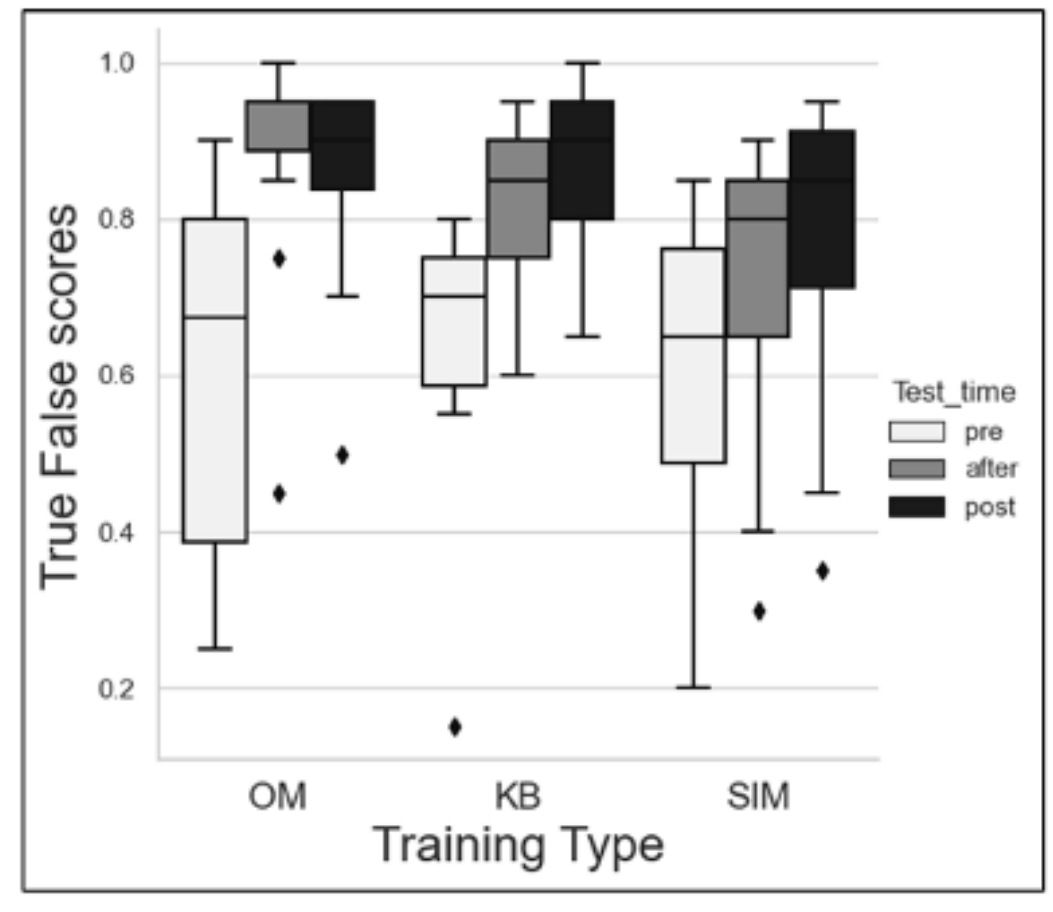

**Figure 6b: True False Score by Training Type**

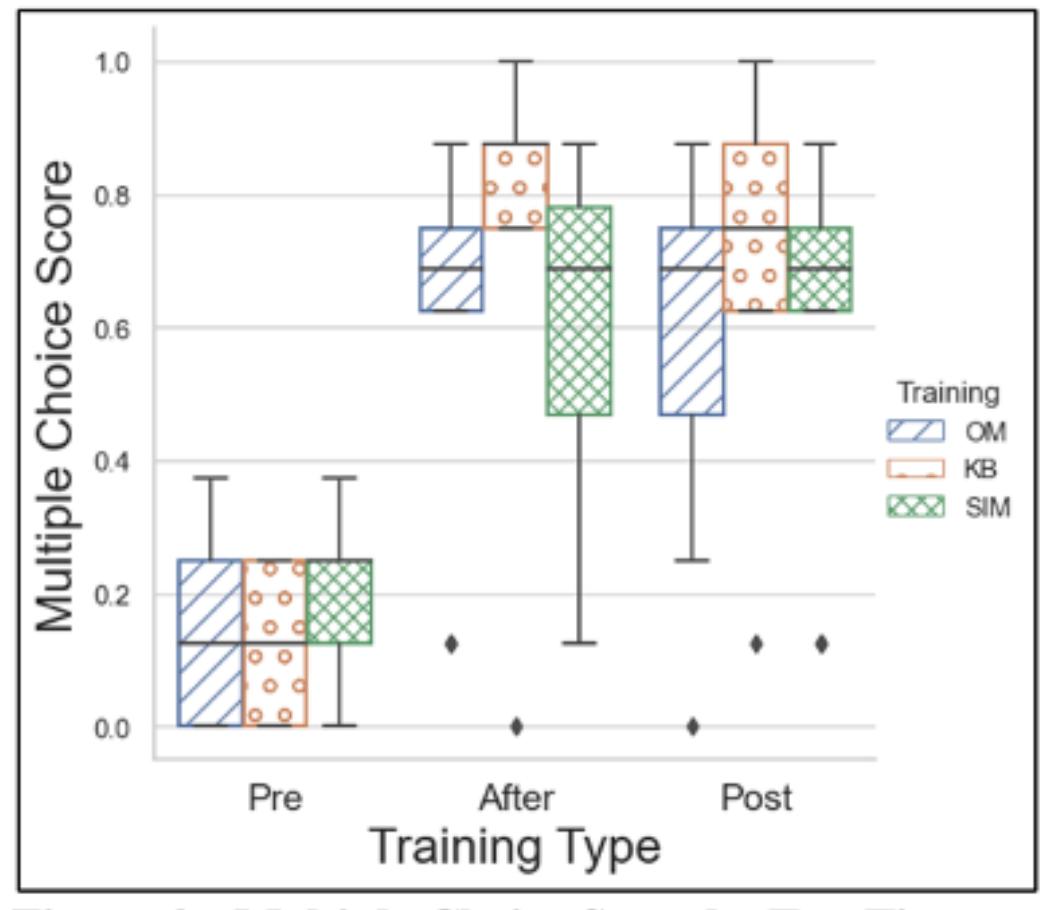

**Figure 6c: Multiple Choice Score by Test Time**

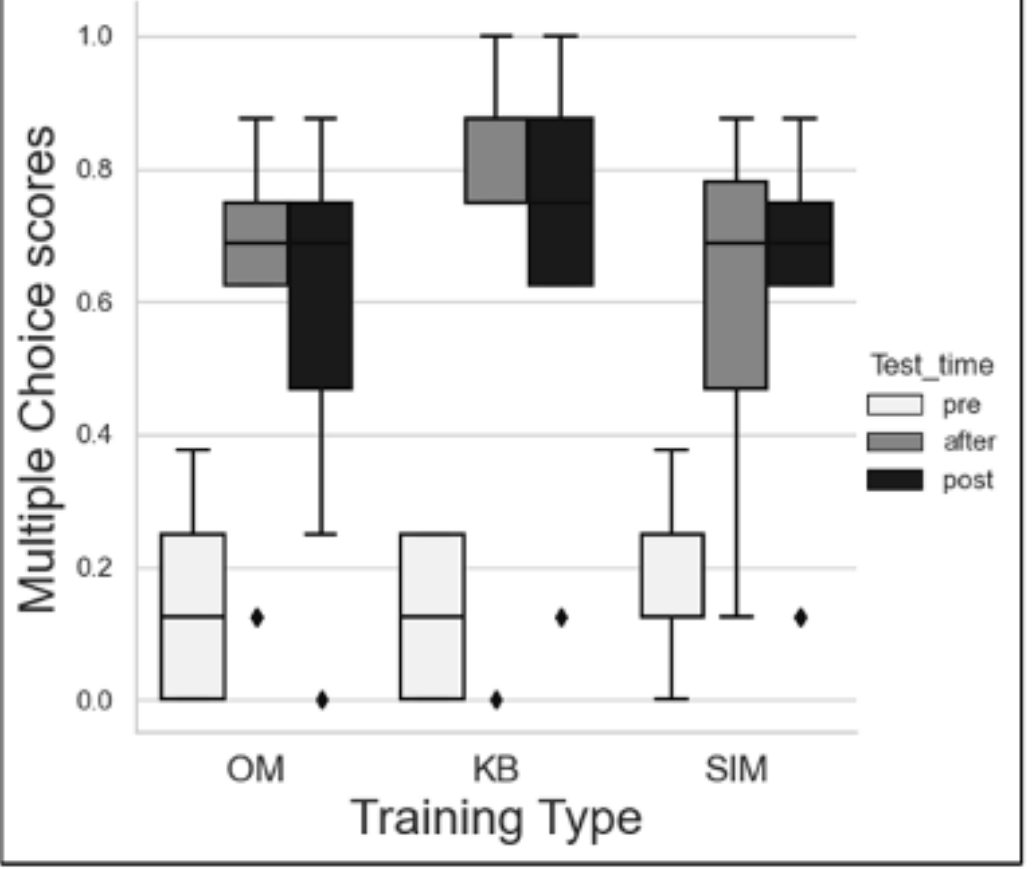

**Figure 6d: Multiple Choice Score by Training Type**

**Figure 6. Score comparison by training group and test time**

(Note: Pre: score before training; After: score right after training; Post: score after one week.
OM: Owners’ manual training, KB: Knowledge-based training, SIM: Skill-based simulator training.)

**Table 4. Summary of Mixed Regression on True/False Scores with Corresponding Cohen's Effect Size (Model 1)**

| Variables | Coef. | S.E. | P-value | Effect size | |
|---|---|---|---|---|---|
| | | | | Cohen's d | 95% C.I. |
| Intercept | 0.64 | 0.06 | < 0.001 | -0.56 † | [-1.15, 0.04] |
| Training [KB] (base: OM) | 0.04 | 0.07 | 0.593 | 0.19 | [-0.52, 0.90] |
| Training [SIM] | -0.00 | 0.07 | 0.953 | -0.02 | [-0.73, 0.69] |
| **Test time [after] (base: pre)** | **0.28** | **0.05** | **< 0.001*** | **1.43 ‡** | **[0.89, 1.96]** |
| **Test time [post]** | **0.25** | **0.05** | **< 0.001*** | **1.28 ‡** | **[0.74, 1.81]** |
| Male (base: female) | -0.02 | 0.04 | 0.653 | -0.1 | [-0.56, 0.35] |
| Older (base: middle) | -0.05 | 0.04 | 0.256 | -0.26 | [-0.72, 0.19] |
| Training [KB]: Test time [after] | -0.10 | 0.04 | 0.203 | -0.49 | [-1.25, 0.27] |
| **Training [SIM]: Test time [after]** | **-0.17** | **0.07** | **0.029 *** | **-0.85 ‡** | **[-1.61, -0.09]** |
| Training [KB]: Test time [post] | -0.04 | 0.04 | 0.617 | -0.19 | [-0.95, 0.57] |
| Training [SIM]: Test time [post] | -0.08 | 0.04 | 0.293 | -0.4 | [-1.16, 0.35]) |

(**Note**: * significant *p*-value, α= .05; † medium effect size, d≥ .5; ‡ large effect size, d≥ .8. Pre: score before training; After: score right after training; Post: score after one week; OM: Owners' manual training, KB: Knowledge-based training, SIM: Skill-based simulator training.)

**Table 5. Summary of Mixed Regression on Multiple-Choice Scores (Model 2)**

| Variables | Coef. | S.E. | P-value | Effect size | |
|---|---|---|---|---|---|
| | | | | Cohen's d | 95% C.I. |
| **Intercept** | **0.17** | **0.08** | **0.030 *** | **-1.00 ‡** | **[-1.46, -0.54]** |
| Training [KB] (base: OM) | -0.01 | 0.09 | 0.907 | -0.03 | [-0.57, 0.51] |
| Training [SIM] | 0.07 | 0.09 | 0.413 | 0.23 | [-0.32, 0.77] |
| **Test time [after] (base: pre)** | **0.49** | **0.06** | **< 0.001 *** | **1.51 ‡** | **[1.13, 1.89]** |
| **Test time [post]** | **0.46** | **0.06** | **< 0.001 *** | **1.42 ‡** | **[1.04, 1.80]** |
| Male (base: female) | 0.03 | 0.06 | 0.670 | 0.08 | [-0.28, 0.44] |
| Older (base: middle) | -0.09 | 0.06 | 0.137 | -0.28 | [-0.64, 0.08] |
| **Training [KB]: Test time [after]** | **0.18** | **0.09** | **0.048** | **0.55 †** | **[0.01, 1.09]** |
| Training [SIM]: Test time [after] | -0.09 | 0.09 | 0.290 | -0.29 | [-0.83, 0.25] |
| Training [KB]: Test time [post] | 0.14 | 0.09 | 0.128 | 0.42 | [-0.12, 0.96] |
| Training [SIM]: Test time [post] | -0.04 | 0.09 | 0.637 | -0.13 | [-0.67, 0.41] |

(**Note**: * significant *p*-value, α= .05; † medium effect size, d≥ .5; ‡ large effect size, d≥ .8. Pre: score before training; After: score right after training; Post: score after one week; OM: Owners' manual training, KB: Knowledge-based training, SIM: Skill-based simulator training.)

As shown in Table 4 and Table 5, age and gender were not significantly associated with questionnaire performance. Consistent with Figure 6, significant increases in true/false scores were observed for the OM group immediately after the training intervention ("after") ($p < .01$; $d = 1.43$) and after the on-road test, about a week later ("post") ($p < .01$; $d = 1.28$). A similar trend regarding the influence of test time was observed for subjects in the KB group and for subjects in the SIM group after the on-road test. A subsequent pairwise comparison (Tukey) indicated that there was no significant difference between the KB and SIM groups. This finding is consistent with the results of Model 1 (Table 4). Additionally, there was no statistically significant difference between test scores immediately after training and those obtained post-training.

Multiple-choice questions showed a significant increase after the training intervention for the OM group ("Test time [after]", $p < .01$; $d = 1.51$) and after the on-road test ("Test time [post]", $p < .01$; $d = 1.42$) (see Table 5). A similar trend was observed for the SIM group and the KB group after the on-road test, as indicated by the insignificant interaction terms. Participants in the KB group scored even higher immediately after the training intervention than subjects in the OM group in the same context. Pairwise comparison ("Tukey") showed no statistically significant difference between the KB and SIM groups for multiple-choice questions.

**LKA Frequency Analysis - Negative Binomial (Model 3)**

As shown in Figure 7, middle-aged subjects in the OM and KB groups appeared to use LAK more frequently than older subjects in the same groups; however, in the SIM group, middle-aged and older subjects performed similarly.

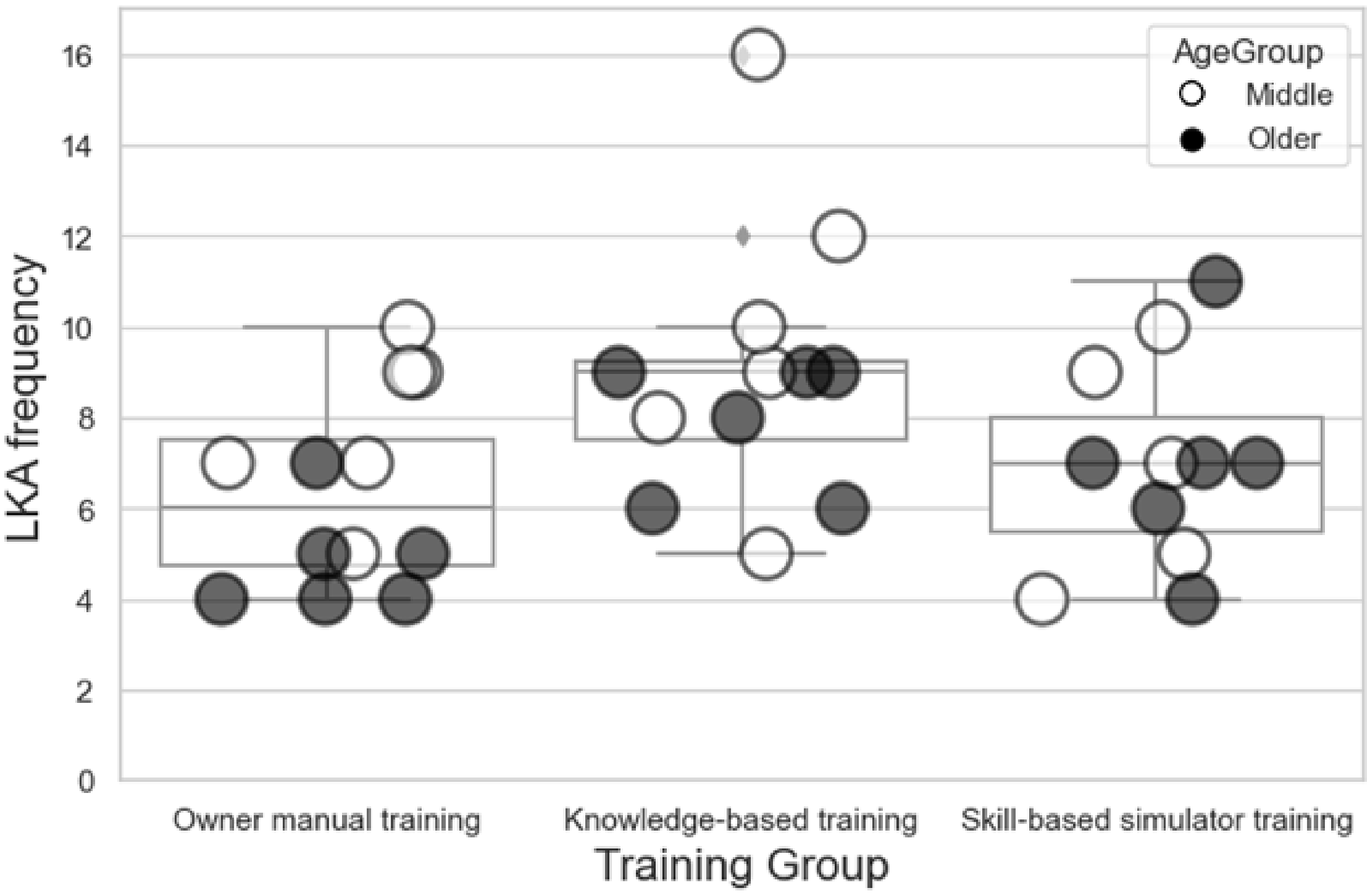

**Figure 7. LKA engagement frequency by training groups with age difference**

**Table 6. Summary of Negative Binomial Model on LKA Engagement Frequency (Model 3)**

| Variables | Coef. | S.E. | P-value | Effect size | |
|---|---|---|---|---|---|
| | | | | Cohen's d | 95% C.I. |

| Intercept | 2.01 | 0.14 | <0.001* | 2.01 ‡ | [1.73, 2.28] |
|---|---|---|---|---|---|
| **Training [KB] (base: OM)** | **0.34** | **0.15** | **0.023 *** | **0.34** | **[0.05, 0.64]** |
| Training [SIM] | 0.11 | 0.16 | 0.516 | 0.11 | [-0.21, 0.42] |
| Older (base: middle) | -0.24 | 0.12 | 0.058 | -0.24 | [-0.48, 0.01] |
| Male (base: female) | -0.11 | 0.12 | 0.382 | -0.11 | [-0.35, 0.13] |

(**Note**: * significant *p*-value, α= .05; † medium effect size, d≥ .5; ‡ large effect size, d≥ .8. Pre: score before training; After: score right after training; Post: score after one week; OM: Owens'r manual training, KB: Knowledge-based training, SIM: Skill-based simulator training.)

A negative binomial model was constructed to investigate the relationship between the variables and the LKA engagement frequency (integers from 1, 2, ...). Initially, interactions between all variables were included in the model, but they were removed as they proved to be insignificant at α = 0.05. As shown in Table 6, the final model without interaction terms found that the LKA engagement frequency for the knowledge-based training group ($p < .05$; $d = 0.34$) is 1.40 times that of the owners' manual training group. A subsequent Tukey pairwise test did not detect a significant difference between the KB and SIM training groups.

### LKA Duration Analysis - Mixed Regression (Model 4)

Through the on-road test and subsequent video coding analysis, the engagement frequencies of the LKA and ACC systems were recorded and counted. Additionally, the continuous active durations of LKA and ACC were calculated. **Figure 8** illustrates the LKA engagement duration in relation to driving time and engagement frequency. The engagement duration decreased again towards the end of the drive. Figure 8a shows the mean engagement duration at each timestamp (in minutes). Since active engagement does not occur every minute, Figure 8a represents the general trend of engagement duration throughout the entire driving test (average driving time = 36.7 minutes).

**Figure 8b** presents the mean duration of each engagement episode, accompanied by 95% confidence intervals. The x-axis indicates individual engagement instances, while the y-axis represents the duration of each active engagement. The results suggest that for both the Owners' Manual (OM) and Simulation-based (SIM) training groups, the duration of LKA engagements was relatively short during the initial instances, followed by a gradual increase over time.

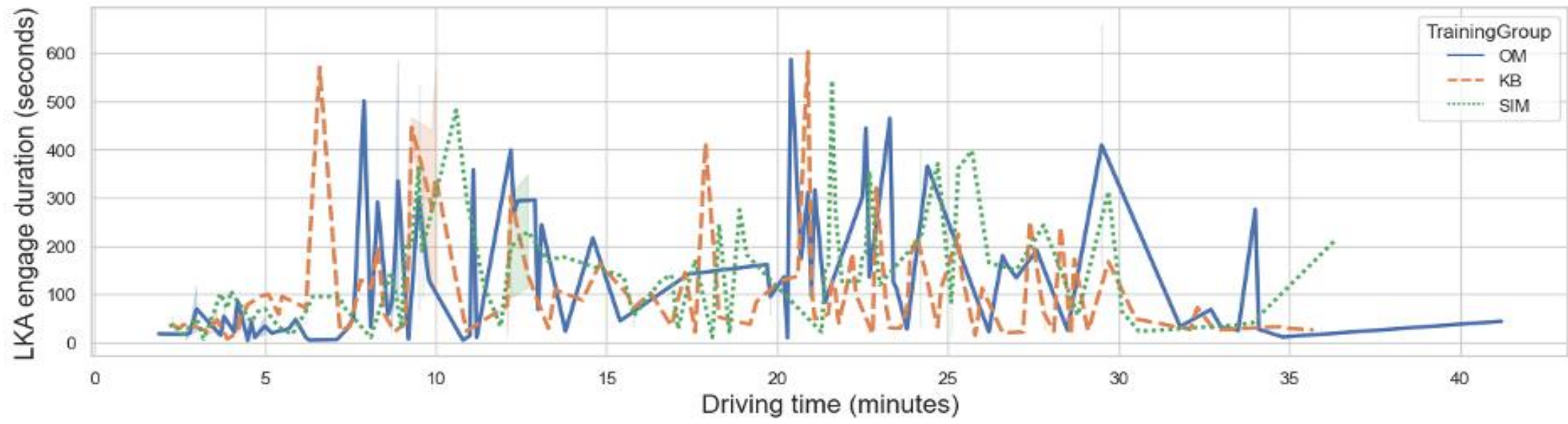

**8a. LKA engage (active) duration through driving time**

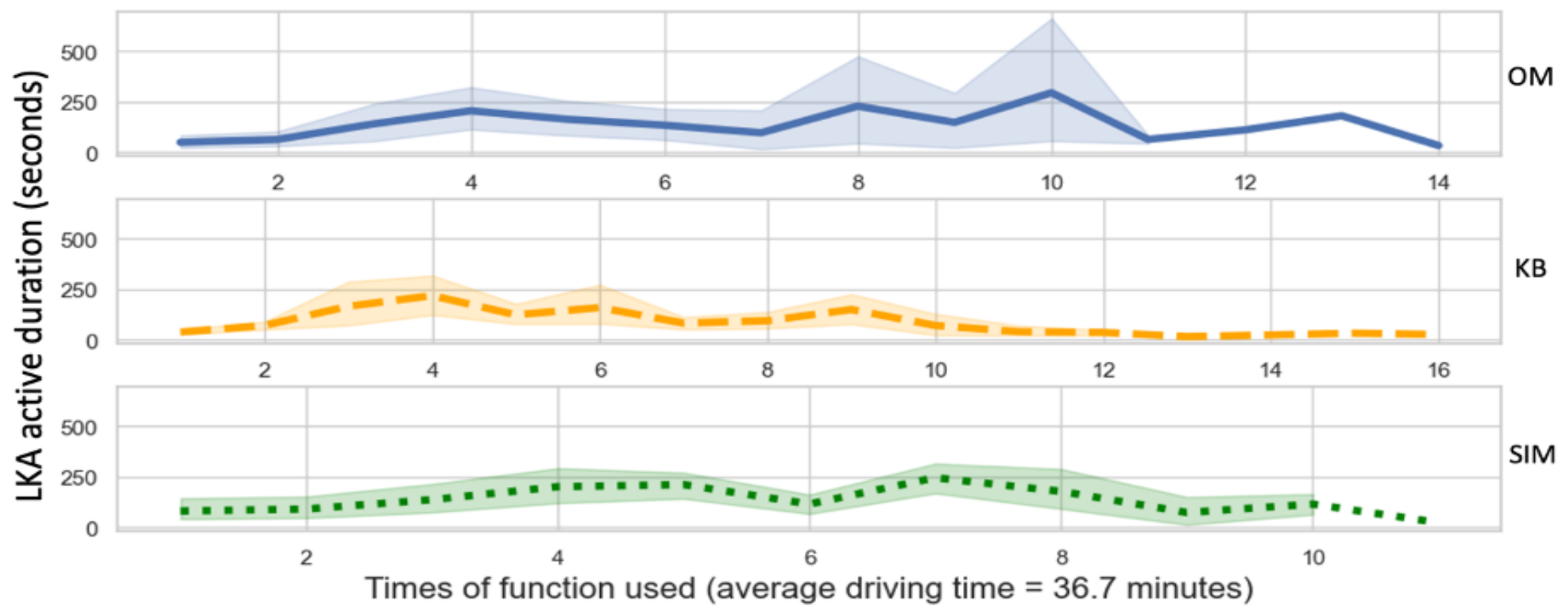

**8b. LKA engage (activation) duration by frequency**

**Figure 8: LKA engage duration during the trip**

(**Note**: Pre: score before training; After: score right after training; Post: score after one week.
OM: Owners' manual training, KB: Knowledge-based training, SIM: Skill-based simulator training.)

### Table 7. Summary of Mixed Regression on LKA Engagement Duration (Model 4)

| Variables | Coef. | S.E. | P-value | Effect size | |
|---|---|---|---|---|---|
| | | | | Cohen's d | 95% C.I. |
| Intercept | 83.06 | 21.47 | <0.001 * | -0.10 | [-0.35, 0.15] |
| Male (base: female) | 10.91 | 16.28 | 0.508 | 0.08 | [-0.16, 0.32] |
| **Older (base: middle)** | **36.01** | **16.35** | **0.036 *** | **0.27** | **[0.03, 0.50]** |
| Training [KB] (base: OM) | -23.79 | 19.23 | 0.228 | -0.18 | [-0.46, 0.10] |
| Training [SIM] (base: OM) | 6.92 | 20.81 | 0.742 | 0.05 | [-0.25, 0.35] |
| **LKA engaged start time** | **0.03** | **0.01** | **0.016 *** | **0.14** | **[0.03, 0.26]** |

(**Note**: * significant *p*-value, $\alpha = .05$; † medium effect size, $d \geq .5$; ‡ large effect size, $d \geq .8$. Pre: score before training; After: score right after training; Post: score after one week; OM: Owners' manual training, KB: Knowledge-based training, SIM: Skill-based simulator training.)

The final model, selected based on the Akaike Information Criterion (AIC), includes no interaction terms. The model results indicated that older subjects (coef. = 36.01; $p < 0.05$; $d = 0.27$) were more likely to engage in LKA for a longer duration compared to the middle-aged group. Consistent with Figure 8, the mixed model suggested that later LKA engagements were associated with longer durations (coef. = 0.03; $p < 0.05$; $d = 0.14$). A subsequent Tukey pairwise test revealed no significant difference between the KB and SIM training groups.

### ACC Frequency Analysis - Negative Binomial (Model 5)

According to **Figure 9**, the average ACC engagement frequency appears higher for the knowledge-based training group; nonetheless, this difference may not be significant. ACC engagement frequencies are also similar between the two age groups, especially for subjects in the KB and SIM groups.

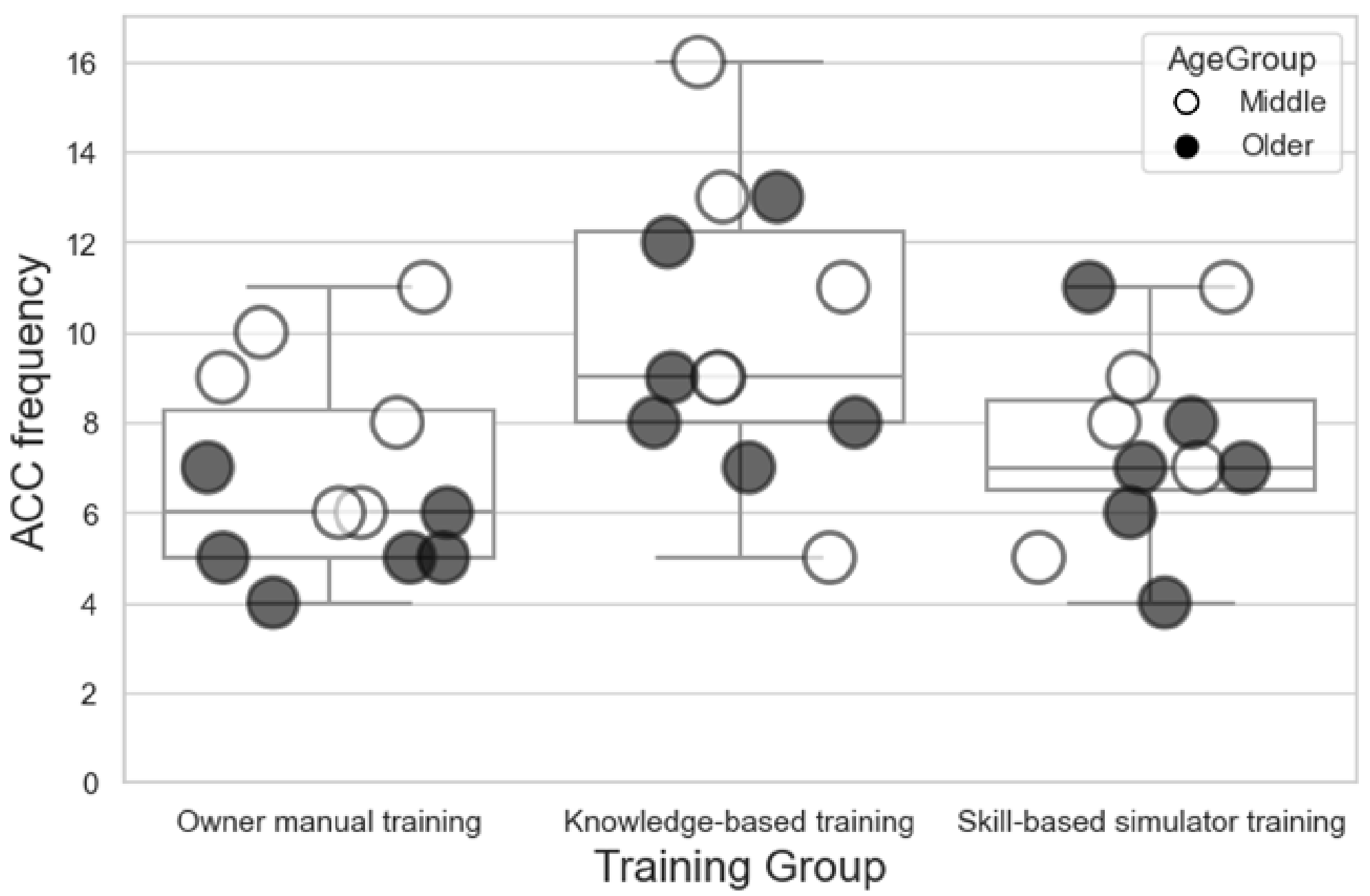

**Figure 9: ACC engagement frequency by training groups with age difference**

**Table 8. Summary of the Negative Binomial Model on ACC Engagement Frequency (Model 5)**

| Variables | Coef. | S.E. | P-value | Effect size | |
|---|---|---|---|---|---|
| | | | | Cohen's d | 95% C.I. |
| Intercept | 2.04 | 0.14 | <0.001 | 2.04 ‡ | [1.77, 2.30] |
| **Training [KB] (base: OM)** | **0.37** | **0.15** | **0.012 *** | **0.37** | **[0.08, 0.66]** |
| Training [SIM] | 0.12 | 0.16 | 0.465 | 0.12 | [-0.19, 0.43] |
| Older (base: middle) | -0.21 | 0.12 | 0.081 | -0.21 | [-0.45, 0.03] |
| Male (base: female) | -0.12 | 0.12 | 0.340 | -0.12 | [-0.36, 0.12] |

(**Note**: * significant *p*-value, α= .05; † medium effect size, d≥ .5; ‡ large effect size, d≥ .8. Pre: score before training; After: score right after training; Post: score after one week; OM: Owners' manual training, KB: Knowledge-based training, SIM: Skill-based simulator training.)

The final model, selected based on the AIC, includes no interaction terms. The results indicate that the ACC engagement frequency for the knowledge-based training group ($p < 0.05$; $d = 0.37$) is 1.45 times that of the owners' manual training group. No significant difference in ACC engagement frequency was observed between the owners' manual group and the skill-based group. Additionally, the Tukey pairwise comparison found no significant difference between the KB and SIM training groups.

### ACC Duration Analysis - Mixed Regression (Model 6)

Similar to LKA engagement duration, Figure 10 separately illustrates the ACC engagement duration by driving time and frequency. Figure 10a shows the mean engagement duration at each timestamp (in minutes). Since active engagement does not occur every minute, Figure 10a provides the general trend of engagement duration throughout the entire driving test (average driving time = 36.7 minutes). Figure 10b presents the mean duration of each engagement with a 95% confidence interval; the labels on the x-axis indicate each engagement instance. The plots show that ACC engagement duration was short during the first five minutes but started to increase after 20 minutes.

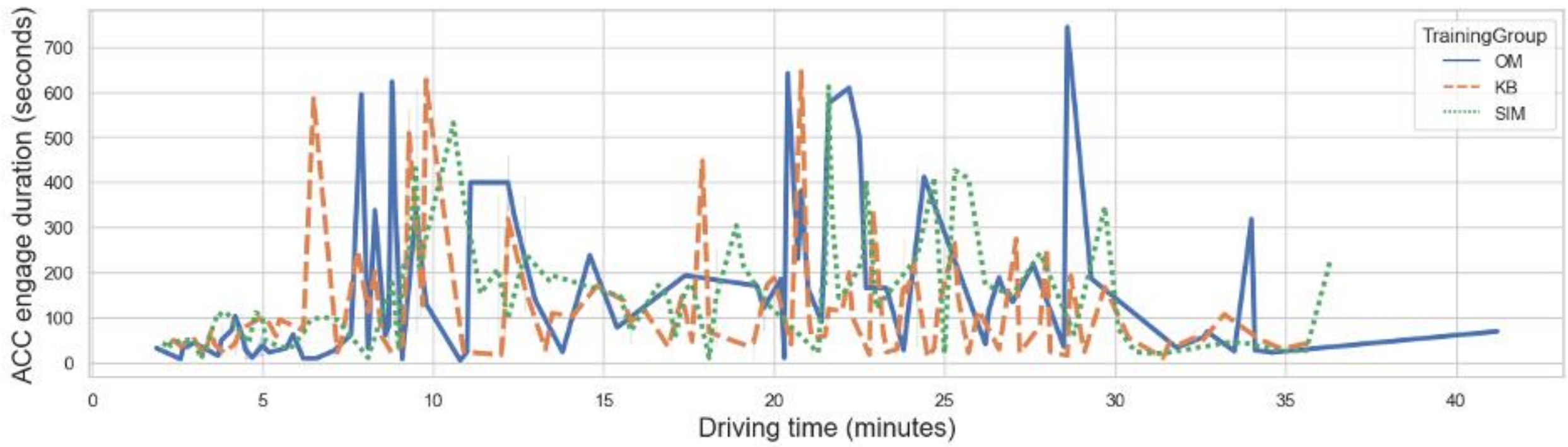

**Fig 10a. ACC engage (active) duration through driving time**

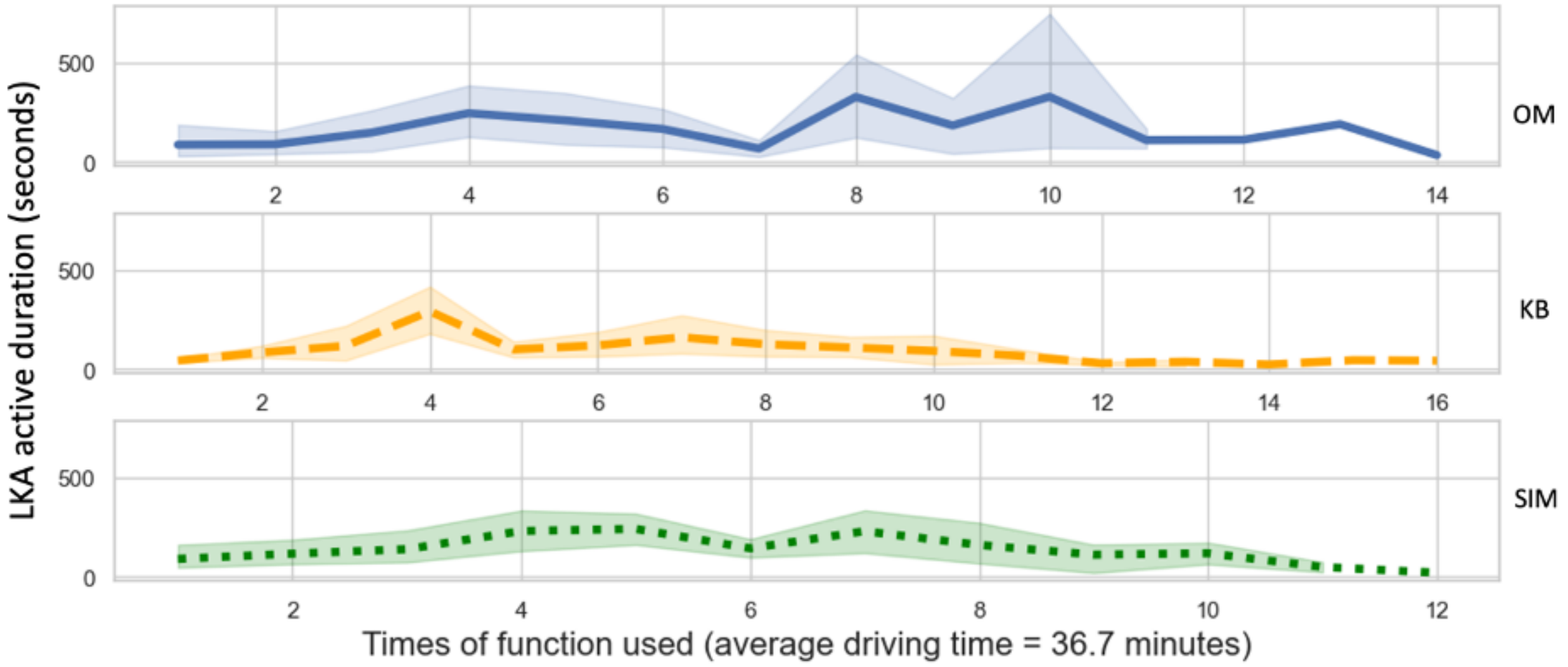

**Fig 10b. ACC engage (activation) duration by frequency**

**Figure 10. ACC engage duration during the trip**

(**Note**: Pre: score before training; After: score right after training; Post: score after one week.
OM: Owners' manual training, KB: Knowledge-based training, SIM: Skill-based simulator training.)

**Table 9. Summary of Mixed Regression on ACC Engagement Duration (Model 6)**

| Variables | Coef. | S.E. | P-value | Effect size | |
|---|---|---|---|---|---|
| | | | | Cohen's d | 95% C.I. |
| Intercept | 89.19 | 29.89 | 0.003 * | 0.04 | [-0.20, 0.28] |
| Male (base: female) | 0.64 | 17.97 | 0.971 | 0.00 | [-0.22, 0.23] |
| **Older (base: middle)** | **45.18** | **18.00** | **0.013 *** | **0.29** | **[0.06, 0.52]** |
| Training [KB] (base: OM) | 24.22 | 39.48 | 0.540 | -0.34 | [-0.61, -0.07] |
| Training [SIM] (base: OM) | 1.81 | 43.44 | 0.967 | -0.09 | [-0.39, 0.20] |
| **ACC engaged start time** | **0.06** | **0.03** | **0.022 *** | **0.23** | **[0.03, 0.43]** |

| Training [KB]: ACC engaged start time | -0.08 | 0.04 | 0.025 * | -0.30 | [-0.57, -0.04] |
|---|---|---|---|---|---|
| Training [SIM]: ACC engaged start time | -0.02 | 0.04 | 0.667 | -0.06 | [-0.36, 0.23] |

(**Note**: * significant *p*-value, α= .05; † medium effect size, d≥ .5; ‡ large effect size, d≥ .8. Pre: score before training; After: score right after training; Post: score after one week; OM: Owners' manual training, KB: Knowledge-based training, SIM: Skill-based simulator training.)

The mixed model results suggested that older subjects (coef. = 45.18; $p < 0.05$; $d = 0.29$) are more likely to have longer ACC engagements compared to the middle-aged group. Additionally, the model indicated that ACC engagement duration is associated with the ACC engagement start time, consistent with driving length. However, while ACC engagement duration increased for the owners' manual training group, subjects in the knowledge-based training group did not engage in ACC more frequently. Furthermore, a Tukey pairwise test did not detect a significant difference between the KB and SIM training groups.

## Discussion

This study was designed to evaluate whether and how training can enhance drivers' safe interaction with automated vehicle systems by improving awareness of their capabilities and limitations. Such awareness is critical for preventing misuse, insufficiently timely intervention, and increased crash risk.

A review of owners' manuals from 12 automakers highlighted substantial variation, including inconsistent terminology, particularly in system naming, differing function descriptions, and limited safety guidance on system feasibility and limitations. Providing clearer details about the driving conditions in which VA systems may not operate reliably would support safer driver decision making and encourage appropriate system use. Addressing these inconsistencies represents an opportunity to strengthen driver understanding, promote realistic expectations, and encourage appropriate reliance on automation. The findings suggest that standardized safety guidelines for automakers' manuals could help ensure that drivers receive consistent, safety-relevant information. Beyond documentation, structured training interventions are also valuable, especially for drivers with no prior experience, to promote correct use and support safe adoption of VA functions.

Two mixed models (Model 1 and Model 2) on score data indicated that training interventions significantly increased response accuracy, strengthening drivers' ability to recognize system boundaries and supporting safer interactions with automation. Knowledge-based training methods had the most positive influence on post-training outcomes, although all three approaches were effective in enhancing comprehension that is essential for preventing misuse and improving driver readiness to intervene when needed. Skill-based training was somewhat less effective than the owners' manual method for true/false questions. Yet, knowledge-based training produced the most significant gains for multiple-choice items, highlighting its value in reinforcing safety relevant decision making during automated driving.

Regarding LKA and ACC function usage frequency, negative binomial models (Model 3) found that the LKA engagement frequency for the knowledge-based (KB) training group was 1.40 times that of the owners' manual (OM) training group. Model 5 found that the ACC engagement frequency for the KB group was 1.46 times that of the OM group. Neither model observed a difference in function usage frequency between the OM and skill-based (SIM) training groups. These findings suggest that knowledge-based training (summarized operational guidelines and visual aids) not only increased function use but also supported more informed and safety relevant engagement, reducing the likelihood of inappropriate reliance or missed opportunities to intervene. The tablet-based text and graph training materials for the KB group may have provided clearer and more accessible information, which in turn helped drivers recognize when automation could be used appropriately and when manual control was necessary, thereby reinforcing safer operational decisions. As Ge et al. (2015) suggested, visual representation and visualization facilitate information processing and knowledge construction, further demonstrating the potential safety benefits of knowledge-based training.

Two mixed regression models (Model 4 and Model 6) showed that LKA and ACC engagement durations increased over driving time. Participants typically engaged briefly in the first five minutes, often testing the functions early in the drive. As driving continued, engagement duration increased, likely due to growing confidence and trust in the VA systems. Although increased trust can encourage more consistent use, it also highlights the need to ensure that reliance on automation remains appropriate and does not compromise safety. These findings align with literature reviews indicating that trust in automation is essential for its use (Lee et al., 1994), and exposure to ACC increases driver trust (Rudin-Brown et al., 2003). From a safety standpoint, longer engagement should be interpreted with caution, as it may improve comfort with automation while also creating potential risks of delayed intervention if system limitations are misunderstood. Training can therefore help balance trust and caution, reinforcing safe use of automation over extended driving periods. Notably, ACC engagement duration increased more significantly for the OM training group, while LKA engagement duration was slightly shorter for the KB training group, which may have been influenced by variables such as road conditions and traffic volume during the test.

Combined analyses of LKA and ACC engagement from this study revealed that older drivers tended to activate these systems less frequently but maintained engagement for longer durations compared to middle-aged drivers. This extended reliance highlights the importance of ensuring that older drivers understand system boundaries so that prolonged engagement supports safety rather than creating risks of overreliance. Older drivers particularly benefited from ACC in maintaining safe following distances, thereby lowering the likelihood of rear-end collisions, and their confidence in using automation functions increased with experience and familiarity through resources such as vehicle manuals (Liang et al., 2020). These findings suggest that age-tailored training could further reinforce the safe use of automation by helping older drivers balance the benefits of extended system use with the need for timely manual intervention.

This study suggests that targeted training, particularly knowledge-based formats, can significantly enhance drivers' understanding of automated vehicle systems, encouraging safer and more confident use, especially among older drivers. Specifically, visually supported materials and standardized information across owners' manuals can help drivers recognize system limits, respond appropriately to failures, and make safer operational decisions. Such measures can reduce misuse, enhance trust in automation, and support the safe integration of advanced vehicle technologies into everyday traffic environments.

## Conclusion

This study was designed to evaluate whether and how targeted training can enhance drivers' safe use of automated vehicle (AV) systems by improving awareness of their capabilities and limitations. Misunderstanding or overreliance on these systems can lead to insufficiently timely intervention, inappropriate reliance, and heightened crash risk. To address these concerns, a set of scientific methods was employed, including a review of automakers' owners' manuals to identify gaps in safety-related guidance, the development of training interventions aimed at reducing misuse, and the design and conduct of controlled experiments to assess their impact on safety-relevant outcomes.

A comprehensive literature review and analysis of 12 owners' manuals revealed variations in the descriptions and instructions related to ACC and LKA systems. These observations indicate the potential value of developing more consistent guidelines to support driver understanding and promote safe and appropriate system usage, thereby reducing driver confusion and minimizing the likelihood of unsafe reliance.

A mixed-subject experiment involving both laboratory and on-road sessions was conducted with 36 participants, all of whom had no prior experience with VA systems. Participants were randomly assigned to one of three 30-minute training groups: (1) original owners' manual, (2) knowledge-based (tablet-based text and graphics), and (3) skill-based (driving simulator). Quiz results indicated that, in this experimental setting, the training interventions improved participants' understanding of VA systems and enhanced their ability to recognize operational boundaries that are essential for safe engagement and timely intervention. All three training methods demonstrated positive effects, with the knowledge-based training group showing the greatest improvement in comprehension scores and the strongest reinforcement of safety-relevant knowledge.

On-road tests with 35 participants analyzed ACC and LKA activation frequencies and durations. Results from negative binomial models showed that the knowledge-based (KB) group, which received summarized text and graphs with visual aids, had significantly higher engagement frequencies for both functions compared to the owners' manual (OM) group. This suggests that concise, safety-relevant information enhanced comprehension, trust, and appropriate reliance, thereby lowering misuse risk. No significant differences were observed between the OM and simulator (SIM) groups in function usage. Mixed regression models showed that engagement durations increased over time: early activations were brief, but durations lengthened as drivers gained confidence. From a safety perspective, this highlights the need for training that ensures growing trust is balanced with caution to prevent overreliance. Combined analyses revealed that older drivers engaged less frequently but for longer periods than middle-aged drivers, indicating distinct reliance patterns and underscoring the need for age-sensitive safety training.

The combined lab and on-road study revealed that senior drivers demonstrated distinct engagement patterns with automation features, favoring longer and more continuous activations. Tailored training and repeated exposure enhanced their confidence and supported safe interaction with automation, showing that accessible instructional approaches can help drivers of all ages benefit from these technologies. These findings highlight the value of designing inclusive training resources and in-vehicle technologies that align system capabilities with users' diverse cognitive and physical needs, ultimately advancing both accessibility and safety in the integration of vehicle automation into everyday driving.

**Practical Applications**

This study underscores the promising role of VA systems and ADAS in enhancing safety and supporting older drivers as well as individuals with needs. Findings show that senior drivers engage with automation differently, often favoring longer and more continuous activations, yet they also demonstrate increased confidence and safer driving behaviors with exposure and training. As aging populations and individuals with mobility or cognitive challenges represent a growing segment of the driving community, the development of intuitive, reliable, and supportive assistance functions becomes increasingly important. Tailored training programs and accessible in-vehicle technologies can foster confidence, encourage safe interaction, and ensure that automation benefits a wide range of users, ultimately promoting clear guidance, reliable operation, and safer road conditions for all users.

**Limitations and Recommendations**

To better evaluate the effects of training methods on drivers' understanding and engagement with VA systems, future studies should consider additional variables such as road conditions, traffic volume, drivers' experience levels, and driving strategies (aggressive or conservative). This study only compared older and middle-aged participants, so future studies should also include younger drivers. SIM training effects may have varied due to participant familiarity or preferences, and younger participants, who are often more accustomed to simulated technologies, may respond differently. Another limitation of this study is that it assessed only immediate comprehension and engagement within one week, leaving open questions regarding longer-term knowledge retention and sustained behavioral change. Additionally, the effectiveness of simulator-based training is highly dependent on the realism of the scenarios used. Future research should examine how well the selected scenarios reflect real-world driving conditions or provide a clear justification for their selection to ensure external validity. Moreover, the absence of a no-training control group can limit our ability to quantify the absolute effect of training compared to no intervention. When feasible, future studies can consider including such a control group to enable a more comprehensive evaluation of absolute training effectiveness. Finally, combined training methods should be evaluated, as a multimedia approach that integrates knowledge-based and skill-based training, for example providing concise functional summaries followed by hands-on simulation, may be more effective than using either approach in isolation.

## Highlights

- A mixed-subject experiment evaluated training outcomes for ACC and LKA using three approaches: (1) original owners' manual (OM), (2) knowledge-based training (KB), and (3) skill-based simulator training (SIM).

- A review of literature and 12 automakers' manuals highlighted the need for standardized, safety-focused guidelines on system terminology, function descriptions, and driver instructions.
- Knowledge-based training (KB), which used summarized text and visual aids, was most effective in improving drivers' comprehension of VA system capabilities and limitations, reinforcing safety-relevant knowledge.
- Even short training interventions significantly improved drivers' understanding and appropriate use of VA systems, demonstrating the link between targeted training, driver confidence, and safer engagement with automation.
- Senior drivers favored longer and more continuous activations but benefited from tailored training and repeated exposure. These findings highlight the importance of accessible in-vehicle technologies and training resources to support safe and inclusive automation adoption.

## References (24)